\documentclass{article}
\usepackage{amsmath,graphicx,amssymb}
\usepackage{color}
\usepackage{float}
\usepackage{multicol}
\usepackage{makecell}
\usepackage{subcaption}
\usepackage{algpseudocode,algorithm,algorithmicx}
\usepackage{amsfonts}
\usepackage{mathtools}
\usepackage{multirow}
\usepackage{bm}
\usepackage{nicefrac}
\usepackage{tabularx}
\usepackage{booktabs}
\usepackage{xcolor}
\usepackage{amsthm}
\usepackage{extarrows}
\usepackage{balance}
\usepackage{subcaption}
\usepackage[backend=bibtex,style=ieee,minbibnames=1,maxbibnames=4]{biblatex}
\usepackage{spconf}
\usepackage[keeplastbox]{flushend}

\def\time{{\mathrm{t}}}
\def\freq{{\mathrm{f}}}
\def\mic{{\mathrm{m}}}

\renewcommand{\arraystretch}{1.1}
\newcommand{\ra}[1]{\renewcommand{\arraystretch}{#1}}

\title{NICE-Beam: Neural Integrated Covariance Estimators\\for Time-Varying Beamformers}

\name{Jonah Casebeer$^\dagger$\thanks{$\dagger$ Work performed while at Reality Labs Research.}, Jacob Donley, Daniel Wong, Buye Xu and Anurag Kumar}

\address{Reality Labs Research, USA}

\begin{document}
\ninept
\maketitle

\begin{abstract}
Estimating a time-varying spatial covariance matrix for a beamforming algorithm is a challenging task, especially for wearable devices, as the algorithm must compensate for time-varying signal statistics due to rapid pose-changes. In this paper, we propose Neural Integrated Covariance Estimators for Beamformers, NICE-Beam. NICE-Beam is a general technique for learning how to estimate time-varying spatial covariance matrices which we apply to joint speech enhancement and dereverberation. It is based on training a neural network module to non-linearly track and leverage scene information across time. We integrate our solution into a beamforming pipeline, which enables simple training, faster than real-time inference, and a variety of test-time adaptation options. We evaluate the proposed model against a suite of baselines in scenes with both stationary and moving microphones. Our results show that the proposed method can outperform a hand-tuned estimator, despite the hand-tuned estimator using oracle source separation knowledge.
\end{abstract}


\section{Introduction}
Beamforming provides a powerful way to leverage the spatial diversity of a microphone array to enhance, separate, dereverberate or otherwise process multi-channel data~\cite{gannot2017consolidated}. Recently, mobile microphone arrays mounted on devices such as headsets, hearing aids, and other wearables have been leveraged to perform these tasks~\cite{doclo2010acoustic, valimaki2015assisted, doclo2015multichannel}. However, these devices present challenging conditions for multi-channel speech processing due to their mobile nature. While adapting to source movement is well studied~\cite{mukai2003robust, golan2010subspace, nikunen2017separation}, compensating for array-movements, such as pose change, is less explored. Successfully modeling such pose-varying and spatially dynamic scenes requires integrating both spatial (e.g. pose) and temporal information. Processing is further complicated by the fact that spatial statistics are dependent on both the pose of the array and the current sound source activity. This implies that a single pose-change renders it difficult to leverage previously collected statistics in a conventional linear framework~(such as buffered or recursively averaged statistics). In this work, we aim to learn a non-linear covariance estimation algorithm that is robust to pose-changes.

In spatially stationary scenes, spatial statistics can be aggregated and estimated across multi-second windows. However, in spatially dynamic scenes, the statistics must be re-estimated in a rapid manner since long windows can span various spatial responses. Developing estimators which can rapidly re-estimate statistics for spatially stationary scenes is an area of active research. These developments include approaches that are more computationally efficient~\cite{barnov2017qrd}, probabilistic~\cite{souden2011integrated, ito2017probabilistic}, and able to incorporate both long and short-term information~\cite{kubo2019mask}. Several recent works have studied the effects of pose-change in spatially dynamic scenes on conventional beamformers. In the case of uniform circular arrays, sound-field interpolation is effective for rotation compensation~\cite{wakabayashi2021rotation}. With more complex arrays, techniques based on spherical harmonics can estimate rotation-invariant covariance representations for uncorrelated noise fields~\cite{moore2018noise}. When the array is not rigid, pose-change may induce a deformation, which can be compensated for~\cite{corey2019motion}. These algorithms often rely on good spectral separation estimates, such as those produced by a voice-activity detector or a ratio-mask estimator.

The growing success of deep neural networks~(DNN) in single-channel spectral separation and speech enhancement has led to several works integrating single- and/or multi- channel enhancers into beamforming pipelines for spatially stationary scenes~\cite{erdogan2016improved, heymann2016neural}. These works leverage a two part procedure where a pre-trained DNN mask-estimation module is inserted into an existing beamforming pipeline and used for spectral estimation. Training modules of a beamforming pipeline in one-step or end-to-end has proved challenging due to the numerical instability of spatial covariance matrix inversion, but has lead to improved speech recognition performance~\cite{zhang2021end}. Moreover, it is also possible to do away with the beamforming structure entirely and learn DNN based multi-channel modules that learn to leverage spatial information implicitly~\cite{luo2019fasnet, tolooshams2020channel, koyama2020exploring}. However, some works have found that it is advantageous to incorporate explicit spectral and spatial estimation steps either in the form of explicit masking and beamforming~\cite{zhang2021adl, xu2021generalized} or more abstractly~\cite{wang2021sequential}. The impact of motion, such as pose change, on these methods is, to the best of the authors' knowledge, unexplored.

We explore training online covariance estimation methods for scenes with pose changes and propose NICE-Beam, a hybrid DNN and Digital Signal Processing~(DSP) system capable of joint enhancement and dereverberation. Inspired by work in learning optimizers~\cite{casebeer2021auto}, we develop a covariance estimator based on a recurrent neural network~(RNN) and use it within a conventional pipeline composed of a single-channel enhancer and a beamformer. We train the estimator in an end-to-end fashion and sidestep previously encountered numerical instability issues by working directly on the inverse of the covariance matrix. We demonstrate our approach on a time-varying-spatialized version of the Microsoft DNS Challenge dataset~\cite{reddy2021interspeech} and show that this approach is robust to pose-changes and surpasses the performance of several conventional methods in terms of perceptual and objective metrics. We also find that our method can be improved by providing additional information to the learned estimator. We note some advantages of this hyrbid DNN-DSP approach by showing that our model is test-time modular, and recovers gracefully from hardware failures. Finally, we confirm the practical feasibility of our technique by testing that it can operate online and faster than real-time with low computational requirements.

\begin{figure*}[!h]
    \centering
    \vspace{-4mm}
    \includegraphics[width=1\linewidth]{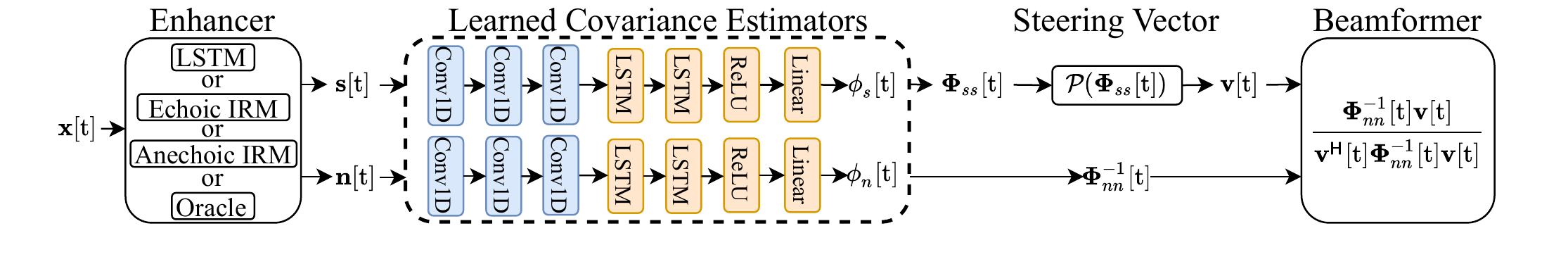}
    \vspace{-9mm}
    \caption{NICE-Beam processing pipeline composed of an enhancer, covariance estimator~(inside the dashed box), and a beamformer. We use separate covariance estimators for speech/noise. The estimator outputs are translated into spatial covariance matrices and fed to the beamformer. In the covariance estimator block, blue denotes an inter-frequency operation whereas yellow denotes an intra-frequency operation.}
    \label{fig:block_diagram}
    \vspace{-4mm}
\end{figure*}

\section{NICE-Beam Formulation}
Let $\mathbf{x}_\mic[\time,\freq] \in \mathbb{C}$ where $\mic \in \{1 \cdots M\}$ be the short-time Fourier transform~(STFT) of a signal at time $\time$ and frequency $\freq$ received by microphone $\mic$ of a moving $M$ channel array. This signal can be decomposed as $\mathbf{x}_\mic[\time,\freq] = \mathbf{d}_\mic[\time,\freq] \ast \mathbf{s}[\time,\freq] + \mathbf{n}_\mic[\time,\freq]$ where $\mathbf{s}[\time,\freq]$ is anechoic speech, $\mathbf{d}_\mic[\time,\freq]$ represents some time-varying impulse response w.r.t microphone $\mic$, $\mathbf{n}_\mic[\time,\freq]$ is noise, and $\ast$ is the convolution operator. The goal of our beamformer is to recover $\mathbf{s}[\time,\freq]$ via a spatial filter $\mathbf{w}[\time,\freq] \in \mathbb{C}^M$ applied to the reverberant mixture~($\mathbf{w}^{\mathsf{H}}[\time,\freq] \cdot \mathbf{x}_\mic[\time,\freq]$). In this work, $(\cdot)^{\mathsf{H}}$ is the Hermitian operator.

The spatial filter, $\mathbf{w}[\time,\freq]$ is estimated using the spatial covariance matrices of the speech~($\mathbf{\Phi}_{ss}[\time,\freq]$) and noise~($\mathbf{\Phi}_{nn}[\time,\freq]$). Typically, it is assumed that the speech/noise covariance matrices are time-invariant and can be computed in a simple running manner using speech estimate $\hat{\mathbf{s}}[\time,\freq]\in \mathbb{C}^M$ and noise estimate $\hat{\mathbf{n}}[\time,\freq]\in \mathbb{C}^M$. However, due to the time-variation of $\mathbf{d}_\mic$, this is not a valid assumption. In this work, we explore using a RNN to estimate speech and noise covariance matrices in an online fashion by nonlineary aggregating $\hat{\mathbf{s}}[\time,\freq]$ and $\hat{\mathbf{n}}[\time,\freq]$. We omit the $\freq$ index when possible.

Filtering in the spatial domain enables constraints which improve beamforming performance. As such, translating the RNN output into something suitable for the computation of a spatial filter is a key contribution of this work. We explore three interpretations of the RNN output and discuss how they correspond to enforcing different degrees of structure. In each of these formulations, we operate on the inverse noise covariance and effectively sidestep previously discovered issues of training RNNs with a beamformer. We call this approach NICE-Beam. Let $g_\theta(\hat{\mathbf{n}}[\time], \mathbf{h}[\time])$ represent the application of a multi-layer RNN parameterized by $\theta$, with $\mathbf{h}[\time] \in \mathbb{R}^D$ as the $D$ dimensional time-frequency dependent state, to an estimate of the noise, $\hat{\mathbf{n}}[\time]$. We omit noise and speech specifiers when possible.

\subsection{Learned Rank-1 Update Estimates}
First, we explore interpreting the network output as a rank-1 symmetric update to an existing $\mathbf{\Phi}^{-1}[\time - 1]$. This is done by configuring $g_\theta(\cdot,  \mathbf{h}[\time])$ to output $\mathbf{\phi}[\time] \in \mathbb{C}^M$ and computing the update as

\begin{equation}
    \mathbf{\Phi}^{-1}[\time] = \mathbf{\phi}[\time]\mathbf{\phi}^{\mathsf{H}}[\time] + \mathbf{\Phi}^{-1}[\time - 1].
\label{eq:rank1}
\end{equation}
The update in \eqref{eq:rank1}, preserves most of the structure of a covariance matrix and is a natural first step towards a learned module. In subsequent sections we call this approach ``Rank-1".

\subsection{Learned Cholesky Estimates}
Second, we allow the network to manage updates internally, in a non-linear fashion. Though, we enforce that the output be a valid inverse covariance by predicting it's Cholesky decomposition. In this formulation, we interpret the network output, $\mathbf{\varphi}[\time] \in \mathbb{C}^{M \times M}$, as a lower triangular matrix within a Cholesky decomposition. To get a lower triangular matrix we zero out elements above the diagonal and pass elements on the diagonal through the absolute value function. The estimate is then given as

\begin{equation}
\mathbf{\Phi}^{-1}[\time] = \bar{\mathbf{\varphi}}[\time]\bar{\mathbf{\varphi}}^{\mathsf{H}}[\time],
\label{eq:cholesky}
\end{equation}
where $\bar{\mathbf{\varphi}}[\time]$ is the lower-triangular absolute-value modified $\mathbf{\varphi}[\time]$. In later sections we call this approach ``Cholesky".

\subsection{Learned Arbitrary Estimates}
Lastly, we explore removing all constraints and allow the RNN to output an arbitrary square matrix, $\mathbf{\xi}[\time] \in \mathbb{C}^{M \times M}$. Specifically, 
\begin{equation}
\mathbf{\Phi}^{-1}[\time] = \mathbf{\xi}[\time]. 
\label{eq:arbitrary}
\end{equation}
We call this approach ``Arbitrary" in subsequent sections as any structure within the estimate is learned and not by construction.

\subsection{Training and Leveraging Learned Estimators}
To train a NICE-Beam estimator, we construct the pipeline from Fig.~\ref{fig:block_diagram}. Inputs are passed through an enhancer, used to estimate a covariance, and then fed to a beamforming module. We compute a loss on the beamformer output, since it is not clear what the optimal covariance estimate is. However, we do know the desired output~(noise-less dry speech). Our formulation produces online estimators, which are custom learned for the training data and metrics. This allows us to train estimators, which specialize in speech separation for wearable arrays of a specific geometry and outperform more generically formulated estimators. Another advantage of this approach is the ability to leverage additional input signals by passing additional inputs to $g_\theta(\cdot)$ besides the current spatial STFT frame and state. We explore providing adjacent frequency information when computing  $\mathbf{\phi}_n[\freq, \time]$. Specifically, we use information from frequencies adjacent to $\hat{\mathbf{n}}[\freq, \time]$ and learn a frequency information sharing network $f_\psi(\hat{\mathbf{n}}[\freq - k:\freq + k, \time])$, parameterized by $\psi$, whose output is passed to $g_\theta(\cdot)$. Since we are training a custom estimator for speech signal enhancement, $f_\psi(\cdot)$ could learn to leverage the spectral structure of speech. The particular selection of enhancer, beamformer, loss function, $g_\theta(\cdot)$, and $f_\psi(\cdot)$ is discussed in the next section.

\section{Experimental Setup}
To evaluate our learned online covariance estimators, we implement the pipeline shown in Fig~\ref{fig:block_diagram}. This pipeline is composed of an enhancer, covariance estimator, and a beamforming module. To isolate analysis of the covariance estimator, we perform initial ablations and experiments using a fixed semi-oracle enhancement module. We later extend our analysis to DNN based enhancers. For all evaluated pipelines, we use the minimum variance distortionless response~(MVDR) beamformer, $512$ point Hann analysis/synthesis window, $256$ point hop size and a sampling rate of {$16$\,kHz}. We train separate models for spatially stationary and dynamic scenes. 

\subsection{Datasets}
We created a spatialized version of the the DNS challenge dataset~\cite{reddy2021interspeech} using pyroomacoustics \cite{scheibler2018pyroomacoustics} for simulations. Each scene lasts $5$\,seconds, contains $1$ to $3$ noise sources, $1$ speech source, and is randomly mixed at SNR between $-5$\,dB and $5$\,dB. We simulate rooms with dimensions uniformly distributed between $4$\,m and $8$\,m, and with a T60 between $0.25$\,s and $0.75$\,s. Within each room, we simulate a $7$\,cm diameter circular microphone array with $6$ microphones. For the dynamic dataset, we simulate array rotation by computing fixed impulse responses at $250$ rotations and perform a time-varying convolution. We generate rotation patterns according to a Markov model derived from the EasyCom dataset~\cite{donley2021easycom}. To prevent leakage we use separate room geometries, rotation patterns, and speech/noise files for train/validation/test. We generated two separate datasets, one spatially static~(no rotations) and one spatially dynamic~(with rotations). The train, validation, and test set splits of each dataset contain $20,000$, $5,000$, and $5,000$ scenes respectively. 

\subsection{Enhancer}
To evaluate the effect of enhancement quality and isolate analysis of the covariance estimator, we experiment with semi-oracle, learned and oracle enhancers. For the semi-oracle enhancers we use an IRM~(ideal-ratio mask) computed using either the echoic~(Echoic IRM) or anechoic~(Anechoic IRM) clean speech. For the learned enhancer, we predict a single-channel real-valued mask using a 3-layer LSTM with hidden size of either $256$ or $512$ trained with SI-SDR~\cite{le2019sdr} loss w.r.t the anechoic clean speech. This is a simple but capable model. All masks are applied via $\mathbf{M} \odot \mathbf{x}_{\mic}[\time]$ where $\mathbf{M}$ is the mask and $\odot$ is the element-wise product. For the Oracle enhancer, we set $\hat{\mathbf{s}}=\mathbf{s}$ and $\hat{\mathbf{n}}=\mathbf{x} - \mathbf{s}$. The IRM and oracle enhancers use ground truth knowledge and only serve as comparison points. The LSTM serves as an example of a suitable model whose performance lies between the Echoic and Anechoic IRMs.

\subsection{Beamformer}
For beamforming, we use MVDR with steering vector based on the estimated relative-transfer function. When running NICE-Beam, we use the normalized first column of the target covariance matrix as a steering vector. We do this to reduce computational complexity during training. When testing all other covariance estimators, we use the normalized largest principal component of the target covariance matrix. We denote the extraction of this time-varying steering vector with the operation $\mathcal{P}(\cdot)$. We then compute MVDR weights as

\begin{equation}
\mathbf{w}[\time] = \frac{\mathbf{\Phi}_{nn}^{-1}[\time]\mathbf{v}[\time]}
{\mathbf{v}^{\mathsf{H}}[\time]\mathbf{\Phi}_{nn}^{-1}[\time]\mathbf{v}[\time]},
\label{eq:mvdr}
\end{equation}
where $\mathbf{v}[\time]  = \mathcal{P}(\mathbf{\Phi}_{ss}[\time])$.

\subsection{Covariance Estimates}
We experiment with two conventional covariance estimators as well as learned methods. On the conventional side, we experiment with offline Fixed and online Buffered estimates, which we pair with the enhancers discussed above. For the Fixed case, we compute a single acausal estimate across the entire scene. For the Buffer case, we use a sliding buffer to compute an online time-varying estimate. In each experiment, we tune the buffer size with a grid search over $[5, 50]$ on the validation dataset. We tested exponentially smoothed updates but found they performed worse and chose not to include them.

The learned estimators, $g_\theta(\cdots)$, are composed of a 2-layer $D=128$ dimension hidden size unidirectional LSTM. We stack the real/imaginary components of each STFT bin as input. On the output, we predict the real and imaginary parameters independently using a set of linear layers. When experimenting with frequency information sharing we set $f_\psi(\cdots)$ to be a stack of three convolutional layers each with kernel size $3$ and $64$ channels. The convolutional layers run across frequency while the LSTMs run on each time/frequency independently effectively performing parameter sharing across all frequencies. This makes the model independent of the STFT frame/hop but constrained to a predetermined number of microphones. We implemented this framework in PyTorch.

\subsection{Training}
We use an identical training recipe for training enhancers and estimators individually and jointly. We use SI-SDR loss for training and early stopping and cease training after $50$ epochs. We use the Adam optimizer with a learning rate of $3\times10^{-4}$, gradient clipping, $16$ bit parameters and set the batch size between $8$ and $16$ depending on model size. We performed tuning on the validation set and did not use the test set until all hyperparameters were finalized. Using one Nvidia P100 GPU, it would take $2$ to $5$ days to train a model.
\section{Results}
We evaluate our approach in spatially stationary and dynamic scenes. For comparison, we construct several validation-set-tuned conventional estimators. We also evaluate the test-time modularity of our approach by modifying portions of the processing pipeline without retraining. In all, we find that NICE-Beam estimators outperform their conventional counterparts by a significant amount in all metrics, are modular, robust, and computationally efficient.

\subsection{Static Scenes}
First, we study static scenes and evaluate all methods using SI-SDR, STOI, and PESQ. For this initial analysis, we equip all learned estimator models with a semi-oracle Echoic IRM as the enhancer. These results are shown in Table~\ref{table:static_results}. The first grouping~(rows 0-3) display the results for a reference microphone and three single-channel models, which do not estimate spatial statistics. The next grouping~(rows 4-7) shows results for a variety of conventional estimators paired with a variety of enhancers. We evaluated all pairs of enhancers/estimators and chose to only show the best. The third grouping~(rows 8-9) shows a NICE-Beam estimator which predicts Rank-1 updates and how incorporating learned information from adjacent frequencies~(+F-Info.) improves performance in all metrics. Based on this, we use learned frequency information sharing in all other NICE-Beam models. The final grouping~(rows 10-11) demonstrates that relaxing the enforced structure of the estimator and shifting all covariance processing within the learned module improves performance. Both the Cholesky and Arbitrary models outperform the conventional estimators equipped with perfect source separation knowledge. This is remarkable since the learned modules are causal and rely on worse speech/noise estimates.

The model in row 4 corresponds to \cite{erdogan2016improved, heymann2016neural} where a DNN enhancer is combined with a conventional covariance estimator. Our proposed method is compatible with this approach as we demonstrate in the next section where we study spatially dynamic scenes.

\subsection{Dynamic Scenes}
\vspace{-.75mm}

Next, we study spatially dynamic scenes. These results are shown in Table~\ref{table:dynamic_results} and can be compared to Table~\ref{table:static_results}. The single-channel approaches in rows 0-3 are minimally impacted by spatial changes since they do not perform spatial processing. The next grouping~(rows 4-7) shows results for several pairs of non-learned estimators and enhancers. Unsurprisingly, the fixed estimators shown in rows 4 and 5 drop in performance from static to dynamic scenes. This occurs since collected statistics are dependent on array pose. The buffer based model shown in row 6 maintains performance as it was re-tuned for dynamic scenes. The second to last grouping~(rows 7-9) shows NICE-Beam estimators with Rank-1, Cholesky and Arbitrary structure. While the Rank-1 estimates~(row 7) do not outperform the models with oracle knowledge~(row 6), the Cholesky~(row 8) and Arbitrary~(row 9) estimates do. Though, they do drop in performance when compared to their results in static scenes. Interestingly, the performance drop is largest for the Rank-1 model. We hypothesize that the fully learned state of the Cholesky and Arbitrary models can adapt more rapidly to motion. Though, sometimes additional structure is useful. The Cholesky model scores marginally better in perceptual metrics than the Arbitrary model, indicating that Cholesky structure could be more faithful to the MVDR objective.

In the final grouping~(rows 10-11) we replace the Echoic IRM, a practically unrealizable enhancer, with an LSTM enhancer. In effect, we create a NICE-Beam model with multiple learned components and extend~\cite{erdogan2016improved, heymann2016neural}. Though, instead of a two part training procedure we train the enhancer and estimator jointly, in an end-to-end fashion. This model is entirely realizable as it is online and uses no ground truth knowledge. The only other comparable models are in rows 2 and 4. To push performance, we double the estimator state size from $128$ to $256$. This largest model achieves our best dynamic scene SI-SDR of $7.13$\,dB. We believe these models could perform better with a more exhaustive hyperparameter search and tuning.

\vspace{-.5mm}
\subsection{Test Time Adaptation}
\vspace{-.75mm}

Our approach has interesting modularity properties which enhance it's real-world usability. Since it is part of a modular pipeline, we can modify modules at test time. We study such modifications in dynamic scenes and show results in Table~\ref{table:modularity_results}. Our initial model shown in row 0~(same as Table~\ref{table:dynamic_results} row 9) is the Arbitrary + F-Info model trained with a $512$ window, $256$ point hop, and Echoic IRM.

First, we experiment with doubling and halving the window and hop size at test time~(rows 1-4). Effectively, we evaluate test-time latency and resolution adaption. Next, we experiment with changing the enhancer at test time. In row 5, the masker is changed from an Echoic IRM to an Anechoic IRM. This is a case of swapping in better enhancers at test-time without retraining. However, the enhancer modification can not be too large. In row 6, we provide oracle separation and all metrics drop. We believe this stems from the model being trained with an IRM based enhancer. Finally, we study robustness and simulate hardware failures by replacing a random microphone signal with white noise~(row 7). As expected, all metrics drop, however, they remain higher than the Echoic-IRM~(Table~\ref{table:dynamic_results} row 1) and buffer estimator, which, in our experiments, errored out.

\vspace{-.5mm}
\subsection{Practical Feasibility}
\vspace{-.75mm}

The estimators have $550$\,K parameters and use $2.5$\,MB of storage. The computational budget is $80$\,MFLOP per frame or $2$\,GFLOP per second. They can run faster than real-time with a real-time-factor of $0.05$ on an Nvidia P100 GPU and $0.9$ on a $2.4$\,GHz Xeon CPU.

\setlength{\tabcolsep}{1.7pt}
\begin{table}[H]
    \centering
    \ra{.7}
    \begin{tabular*}{.99\linewidth}{@{\extracolsep{\fill}}l c c c c c c c@{}}\toprule 
        \# && \multicolumn{2}{c}{Model Attributes} && \multicolumn{3}{c}{Metrics}\\ 
        \cmidrule{3-4} \cmidrule{6-8} && \textit{Enhancer} & \textit{Estimator} && \textit{SI-SDR} & \textit{STOI} & \textit{PESQ} \vspace{1mm}\\
        0 && N/A & N/A &&  -1.10 & .738 & 1.17\\
        1 && Echoic IRM & N/A && -0.06 & .813 & 1.51\\
        2 && LSTM~($512$) & N/A  && 2.32 & .804 & 1.44\\
        3 && Anechoic IRM & N/A && 3.28 & .950 & 3.08 \vspace{-.5mm}\\
        \bottomrule \vspace{-2mm}\\
        4 && LSTM~($512$)& Fixed~\cite{erdogan2016improved, heymann2016neural}  && 3.26 & .855 & 1.43\\
        5 && Anechoic IRM & Fixed   && 3.32 & .866 & 1.50\\
        6 && Oracle & Fixed  && 5.12 & .895 & 1.75\\
        7 && Oracle & Buffer && 4.78 & .898 & 1.75 \vspace{-.5mm}\\
        \bottomrule \vspace{-2mm}\\
        8 && Echoic IRM & Rank-1   && 5.78 & .866 & 1.61\\
        9 && Echoic IRM & Rank-1 + F-Info. && 6.57& .876 & 1.64 \vspace{-.5mm}\\
        \bottomrule \vspace{-2mm}\\
        10 && Echoic IRM & Cholesky + F-Info. &&  7.32& .918 & \textbf{2.26}\\
        11 && Echoic IRM & Arbitrary + F-Info.  && \textbf{7.60} & \textbf{.919} & 2.24 \vspace{-.5mm}\\
        \bottomrule \vspace{-2mm}\\
    \end{tabular*}
    \vspace{-2mm}
    \caption{Results in spatially stationary scenes. Models are made of an enhancer and an estimator. The learned estimators in rows 8-11 outperform their fixed counterparts. The highest scores are boldface.}
    \vspace{-3mm}
    \label{table:static_results}
\end{table}

\setlength{\tabcolsep}{1.7pt}
\begin{table}[H]
    \centering
    \ra{.7}
    \begin{tabular*}{.99\linewidth}{@{\extracolsep{\fill}}l c c c c c c c@{}}\toprule
        \# && \multicolumn{2}{c}{Model Attributes} && \multicolumn{3}{c}{Metrics}\\ 
        \cmidrule{3-4} \cmidrule{6-8} && \textit{Enhancer} & \textit{Estimator} && \textit{SI-SDR} & \textit{STOI} & \textit{PESQ} \vspace{1mm}\\
        0 && N/A & N/A && -1.09 & .738 & 1.17\\
        1 && Echoic IRM & N/A && -0.04 & .812 & 1.51\\
        2 && LSTM~($512$) & N/A  && 2.31 & .804 & 1.43\\
        3 && Anechoic IRM & N/A && 3.34 & .950 & 3.09 \vspace{-.5mm}\\
        \bottomrule \vspace{-2mm}\\
        4 && LSTM~($512$) & Fixed~\cite{erdogan2016improved, heymann2016neural} && 1.43 & .798 & 1.31\\
        5 && Oracle & Fixed  && 1.37 & .820 & 1.42\\
        6 && Oracle & Buffer && 4.04 & .889 & 1.70 \vspace{-.5mm}\\
        \bottomrule \vspace{-2mm}\\
        7 && Echoic IRM & Rank-1 + F-Info. && 4.53& .856 & 1.61\\
        8 && Echoic IRM & Cholesky + F-Info. && 6.60 & \textbf{.916} & \textbf{2.26}\\
        9 && Echoic IRM & Arbitrary + F-Info. && 6.80 & .915 & 2.19 \vspace{-.5mm}\\
        \bottomrule \vspace{-2mm}\\
        10 && LSTM~($256$) & Arbitrary + F-Info. && 6.58 & .865 & 1.75\\
        11 && LSTM~($256$) & Arbitrary~($256$) + F-Info. && \textbf{7.13} & .873 & 1.84 \vspace{-.5mm}\\
        \bottomrule \vspace{-2mm}\\
        \end{tabular*}
    \vspace{-2mm}
    \caption{Results in spatially dynamic scenes. Pairing learned estimators and enhancers yields the best SI-SDR as shown in row 11.}
    \vspace{-3mm}
    \label{table:dynamic_results}
\end{table}

\setlength{\tabcolsep}{1.7pt}
\begin{table}[H]
    \centering
    \ra{.7}
    \begin{tabular*}{.99\linewidth}{@{\extracolsep{\fill}}l c c c c c c@{}}\toprule
        \# && Test-Time Modification && \multicolumn{3}{c}{Metrics}\\ 
        \cmidrule{2-4}\cmidrule{5-7} &&&& \textit{SI-SDR} & \textit{STOI} & \textit{PESQ} \vspace{1mm}\\
        0 && No Modification && 6.80 & .915 & 2.19 \vspace{-.5mm}\\
        \bottomrule \vspace{-2mm}\\
        1 && Double Window  && 6.92 & .923 & 2.26\\
        2 && Halve Hop && 6.39 & .910 & 2.05\\
        3 && Double Window and Hop && 6.23 & .911 & 2.16\\
        4 && Halve Window and Hop && 5.11 & .886 & 1.87 \vspace{-.5mm}\\
        \bottomrule \vspace{-2mm}\\
        5 && Anechoic IRM && 4.15 & .916 & 2.21\\
        6 && Oracle Separation && 3.65 & .909 & 2.14 \vspace{-.5mm}\\
        \bottomrule \vspace{-2mm}\\
        7 && Dead Microphone && 0.13 & .832 & 1.48 \vspace{-.5mm}\\
        \bottomrule \vspace{-2mm}\\
    \end{tabular*}
    \vspace{-2mm}
    \caption{Results evaluating test-time modularity. All adaptation is tested on the Arbitrary + F-Info model using the Echoic-IRM.}
    \label{table:modularity_results}
\end{table}

\section{Conclusion}
In this paper, we proposed NICE-Beam, a learning based method for estimating time-varying spatial covariance matrices, with application to joint speech enhancement and dereverberation. We applied these methods on a dataset containing both spatially -stationary and -dynamic scenes. Our empirical results showed that the proposed methods outperform a variety of conventional estimation approaches. Moreover, by incorporating the learned modules into a conventional beamforming pipeline, we can perform real-time inference, and apply test-time adaptation with respect to latency, resolution, and masking modification. The model is also robust to hardware failures. We believe our proposed method could also be a valuable tool for complex spatial processing in other domains.

\newpage
\setlength\bibitemsep{0.4em}
\atColsBreak{\vskip0.4em}
\printbibliography

\end{document}